\documentclass[sn-basic]{sn-jnl}

\usepackage{color}
\usepackage{grffile}
\usepackage{ifthen}
\usepackage{appendix}

\usepackage{graphicx}%
\usepackage{multirow}%
\usepackage{amsmath,amssymb,amsfonts}%
\usepackage{amsthm}%
\usepackage{mathrsfs}%
\usepackage[title]{appendix}%
\usepackage{xcolor}%
\usepackage{textcomp}%
\usepackage{manyfoot}%
\usepackage{booktabs}%
\usepackage{aas_macros}

\begin{document}
	
\title[Structure of Alfvén vortices in planetary magnetospheres]
{On the linear structure of the interlaced Alfvén vortices in the tail of Uranus at solstice.}
\author{\fnm{Filippo} \sur{Pantellini}}\email{Filippo.Pantellini@obspm.fr}
\affil{\orgdiv{LESIA}, \orgname{Observatoire de Paris, Université PSL, CNRS, Sorbonne Université, Université  Paris Cité}, \orgaddress{\street{5 place Jules Janssen}, \city{Meudon}, \postcode{92195}, \country{France}}}

\abstract{Incompressible vortex flow are observed in a large variety of astrophysical plasmas such as the convection zone and the atmosphere of stars, in astrophysical jets in stellar winds and in planetary magnetospheres. 
More specifically, magnetohydrodynamic (MHD) simulations have shown that two large scale interlaced Alfvénic vortices 
structure the magnetic tail of Uranus at solstice time. Assuming identical vortices, we compute the general linear structure of the flow near their centers within the frame of ideal MHD. We then use the analytic results to interpret and qualify the vortices observed in a 3D MHD simulation of a fast rotating Uranus-type planet.      
}	
	
\keywords{Magnetohydrodynamics (1964), Uranus (1751), Planetary magnetospheres (997), Alfven waves (23), Magnetohydrodynamical simulations (1966)}

\maketitle

\section{Introduction}

\begin{figure}
	\center{\rotatebox{0}{\resizebox{0.5\textwidth}{!}
			{\includegraphics{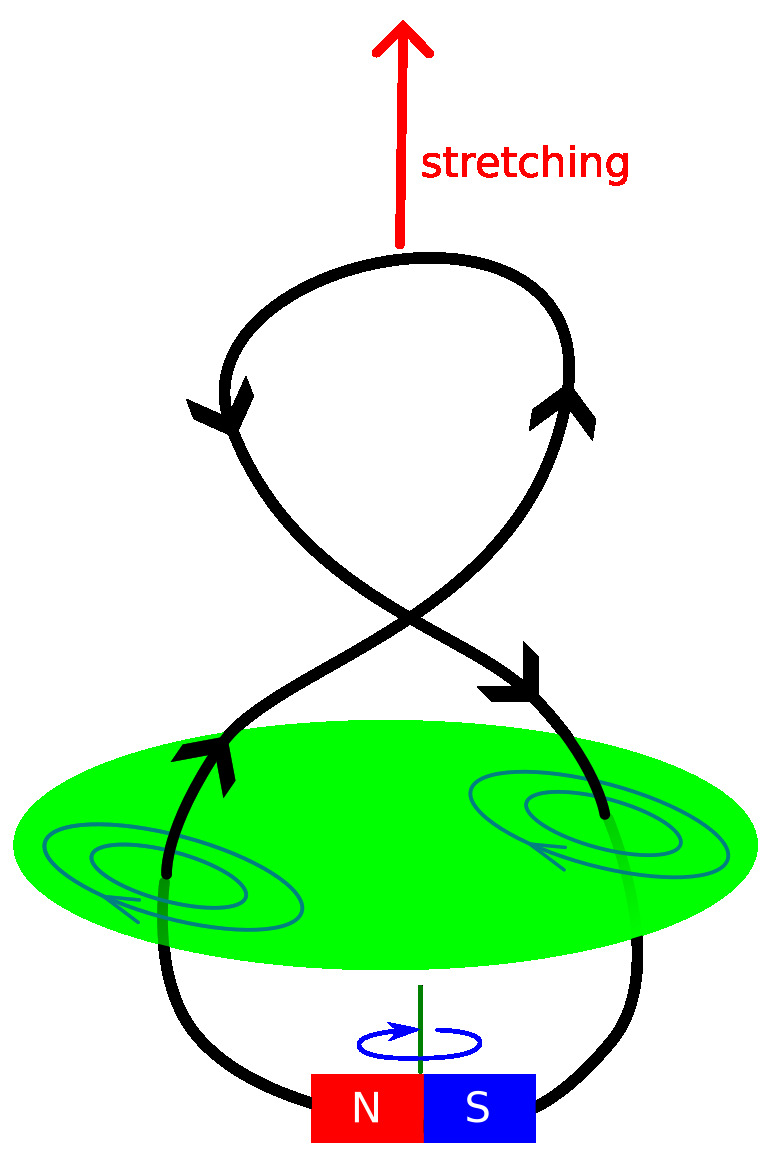}
	}}}
	\caption{Formation of two interlaced Alvénic vortices resulting from the stretching of the field lines from a rotating magnetic dipole. In this particular case (the reference case throughout the paper) the stretching direction is aligned on the rotation axis and perpendicular to the dipole so that the two vortices are identical. The plotted field line is the one which marks the centers of the two vortices in a selected plane oriented perpendicularly to the stretching direction.  
	Also shown are possible trajectories of the footprints of nearby field lines.        
} 
	\label{fig_expanding_dipole}
\end{figure}

Vortices naturally form in moderately to high Reynolds number hydrodynamic flows with velocity shears, making them a common feature in turbulent flows. In complex systems such as the stratified atmosphere of rotating planets, depending on the driving mechanism, they cover a wide range of transverse spatial scales \citep[e.g.][]{Fujita_1981}.   
As already noted by  Helmholtz \citep{Helmholtz_1858}, the vast majority of vortical flows are incompressible and satisfy $\nabla\cdot {\bf u}=0$, where $\bf u$ is the flow velocity vector. In stratified media, they constitute an efficient means for the vertical transport of mass and energy, as in bathtub vortices \citep{Andersen_etal_2003} or in atmospheric tornadoes \citep{WedemeyerBoehm_etal_2012,Kuniyoshi_etal_2023}. In rotating systems, vortices naturally form under the action of the Coriolis force at scales of the order of $u/\Omega$, where  $u$ is a characteristic flow speed and $\Omega$ the angular velocity of rotation \citep{Brito_etal_1995}.  Coriolis driven vortices are often generated in planetary atmospheres \citep{Chan_2005} and in the convective zone of planets and stars \citep{MieschMark_2005}. Hydrodynamic vortices are generally treated theoretically as isolated axisymmetric structures, with the vortex axis representing a straight line \citep{Nakamura_etal_2001,Klimenko_2014}. In some situations, however, the vortex axis can be twisted, not as the consequence of some instability but because 
the vortex is not an isolated one, as for 
the pair of vortices in the paradigmatic example of Figure \ref{fig_expanding_dipole} where a couple of interlaced vortices is formed through the action of stretching the field lines from a rotating dipole. The necessary condition for the winding of the two vortices to occur is that some parts of the field line does not corotate with the footpoints attached to the magnetic dipole. 
In a low $\beta$ plasma ($\beta$ is the thermal pressure to magnetic pressure ratio) significant field line winding requires the length of the field line to exceed the  distance an Alfvén wave can travel during one rotation period of the dipole. Considering a field line of characteristic size $D$ (could be the distance between the footprints on the plane in Figure \ref{fig_expanding_dipole}), the condition reduces to $v_{\rm A} \lesssim \Omega D$, where $v_{\rm A}$ is the Alfvén speed. The condition is generally not met in the solar atmosphere as the Sun is a slow rotator so that for a typical magnetic loop of size $D\sim 5\,10^4\, {\rm km}$ one has $\Omega D\simeq 120 { \,\rm m/s}$ which is much smaller than the typical values of the Alfvén speed in the solar photosphere and above \citep{Tziotziou_etal_2023}.  
For rotating planets, with a significant internal dipole, rough estimates of $\Omega D$ can be obtained by posing $D$ to be of the order of the sub-solar distance of the magnetopause from the planet. Excluding the case of extremely slowly rotating Mercury, $D$ ranges from 10 planetary radii for the Earth to 40 planetary radii for Jupiter (20 for Saturn and 25 for Uranus and Neptune). The resulting values of $\Omega D$ range from $30\,{\rm km/s}$ for the Earth to 
$3\,10^3\,{\rm km/s}$ for Jupiter. These values are either comparable or much larger than the  reference Alfvén speed of $50\,{\rm km/s}$ in  the solar wind surrounding the planets, suggesting that vortices can form in the magnetosphere of these planets. However, for a pair of interlaced vortices of opposite polarity to develop as in the example of Figure \ref{fig_expanding_dipole}, the angle between the rotation axis and stretching direction (i.e. the wind direction) must be smaller than the angle between the rotation axis and the planetary dipole. In the solar system, 
the only planet which meets this requirement is Uranus at solstice, when its rotation axis is approximately aligned with the planet-Sun direction. On this occasion, two large scale interlaced vortices with opposite magnetic polarity form tailwards of the planet \citep{Toth_etal_2004,Griton_etal_2018}. In the magnetic tail of the other rotating planets, only isolated monopolar vortices are expected to form, as the ones observed in simulations of the distant tail of  Jupiter \citep{Zhang_etal_2021a}. 
Isolated vortices also form in astrophysical jets \citep{McKinney_Blandford_2009}, in the solar atmosphere \citep{Wedemeyer_Steiner_2014,Requerey_etal_2018,Battaglia_etal_2021} or in the near-surface convective region of the Sun \citep{Moll_etal_2011} but are often unstable, for example with respect to the current driven kink instability  \citep{Nakamura_Meier_2004,McKinney_Blandford_2009,Wang_etal_2017,Appl_etal_2000} or with respect to the Kelvin-Helmholtz instability \citep{Zaqarashvili_etal_2014,Zaqarashvili_etal_2015}.

\begin{figure}
	\center{\rotatebox{0}{\resizebox{0.5\textwidth}{!}
			{\includegraphics{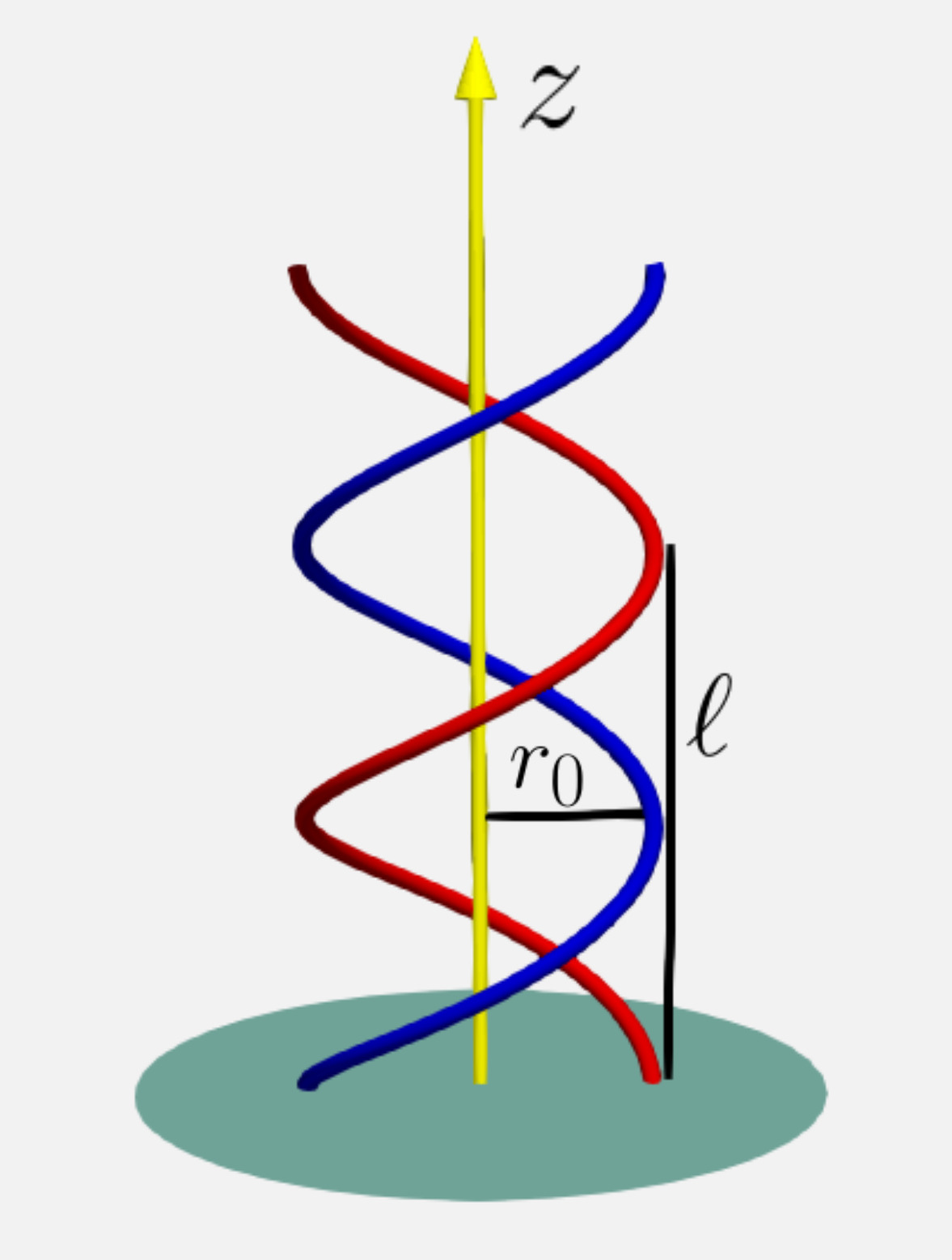}
	}}}
	\caption{Model assumption for the double vortex system. 
		The vortices wind around the $z$ axis at a fixed sistance $r_0$ with a constant pitch $d\varphi/dz=k$, where $\varphi$ is the counterclockwise rotation angle with respect to the $z$ axis. The wavelength $\ell$ defines the wave vector  $|k|=2\pi/\ell$. $k$ is positive/negative for right/left-handed winding.} 
	\label{fig_vortex_geometry}
\end{figure}

On the other hand, two infinitely long, interlaced  identical vortices with opposite magnetic polarity as the ones in Figure \ref{fig_vortex_geometry} 
tend to settle into a stable state of minimum energy  \citep[see][]{Bernstein_etal_1958}. In this paper we assume that such a system of interlaced vortices is a fair representatioWen of the magnetic tail of an idealized Uranus where the planetary field lines are stretched and twisted  as in the schematic of Figure \ref{fig_expanding_dipole}. 
One may argue that the separation between the rotation axis (the $z$ axis) and the magnetopause is not constant when moving tailwards of the planet, implying that the distance between the vortices is also not constant.  
One may also argue that the surrounding solar wind, which flows parallel to the $z$ axis carries a magnetic field, which at Uranus is generally transverse to the $z$ axis and effectively breaks the $\varphi$ isotropy assumed in Figure \ref{fig_vortex_geometry}. 
Concerning the first objection, numerical simulations show that except near the planet, where the magnetic field rotates solidly, the distance of the magnetopause from the rotation axis is small over distances of the order of the winding scale length \citep{Toth_etal_2004}. Concerning the second objection, numerical simulations of a Uranus-type magnetosphere in a magnetized wind by  \cite{Griton_etal_2018} show that the vortices are essentially unaffected by the process of magnetic reconnection between planetary field lines and the field lines carried by the wind.  
With these considerations in mind, it is reasonable to assume that at some distance, tailwards from the planet, there are two interlaced vortices which wind around the $z$ axis at a fixed distance $r_0$ and with a constant pitch $d\varphi/dz=k$ as shown in Figure \ref{fig_vortex_geometry}. Accordingly, we adopt the configuration of Figure \ref{fig_vortex_geometry} for the analytic model presented in this paper. 

The paper is organized as follows: the fluid equations are presented in Section \ref{sec_equations} and the adopted  symmetries in Section \ref{sec_symmetries}. In Section \ref{sec_characteristic} we derive the characteristic form of the fluid equations. The general Alfvénic solution is extensively commented in Section \ref{sec_alfven_mode} and applied to the vortex from a large scale 3D MHD Uranus-type planet simulation in Section \ref{sec_uranus}. Conclusions are given in Section \ref{sec_conclusions}.

\section{The fluid equations in the rotating frame\label{sec_equations}}
We consider incompressible flows at constant plasma density $\rho=\text{const}$ so that the continuity equation reduces to 
\begin{eqnarray}
	\nabla\cdot (\rho {\bf u}) &=& \rho \nabla \cdot {\bf u} + 
	                            {\bf u}\cdot \nabla \rho\\
	                           &=& \rho \nabla \cdot {\bf u}
	                           =0.\label{eq_continuity}
\end{eqnarray}
Sticking to ideal MHD and with the system of interlaced vortices of Figure \ref{fig_vortex_geometry} in mind, 
we write the induction and momentum equations for a neutral conductor with no electric and magnetic polarization in the frame which rotates with the vortices. The system is steady in this frame so that we can drop the temporal variations and write the equations without the $\partial/\partial t$ terms (in units such that $\rho=1$ and the vacuum permeability is $\mu_0=1$): 
\begin{eqnarray}
	0&=&\nabla\times ({\bf u}\times {\bf B})=
{\bf B}\cdot\nabla {\bf u}-{\bf u}\cdot\nabla {\bf B}\label{eq_induction}\\
	{\bf u}\cdot\nabla {\bf u} &=& -\nabla B^2/2 + {\bf B}\cdot\nabla {\bf B} + 2 {\bf u}\times {\bf \Omega} +{\bf r}\Omega^2\label{eq_momentum}
\end{eqnarray}
where $\bf r$ is the vector from the rotation axis (the $z$ axis) to the observation point and ${\bf \Omega}=\Omega\,\hat{\bf z}$ the angular velocity vector (where $\hat{\bf z}$ is the unit vector along the $z$ axis). 
In writing (\ref{eq_induction}) we did use  (\ref{eq_continuity}) and the Maxwell-Gauss law: 
\begin{equation}
	\nabla\cdot {\bf B}=0\label{eq_maxwell_gauss} 
\end{equation}
while in writing (\ref{eq_momentum}) we adopted a polytropic approximation $p\propto \rho^\gamma$ so that, given $\rho=\text{const}$, we have $\nabla p=0$. As first pointed out by \cite{Schiff_1939}, the form of Maxwell's equations in a non-inertial frame differs from their form in an inertial frame, the latter having supposedly be used to construct the system of equations (\ref{eq_induction})-(\ref{eq_maxwell_gauss}). Before a more in depth discussion, we anticipate that the system (\ref{eq_induction})-(\ref{eq_maxwell_gauss}) is appropriate for a charge neutral rotating ideal plasma, as long as the rotation speed (with respect to the inertial frame) of all points of interest is small compared to the speed of light. 
In the case of Uranus, the vortices are confined to within a distance from the rotation not exceeding $100\:R_{\rm U}$,   
so that, given the planet's sidereal rotation period of $17.24\:{\rm h}$, the velocity of a point in the rotating frame is at most of the order of $100\:R_{\rm U} \times 2\pi/17.24\:{\rm h} \simeq 257\:{\rm km/s}$ which is much less than the speed of light (even assuming a ten times faster rotation as in the example of Section \ref{sec_uranus}). Under such circumstances \citep[see e.g.][]{Webster_Whitten_1973}, the magnetic field $\bf B$, the current density $\bf j$ and the effective charge density $\varrho$ are the same in the rotating and in the inertial frame. The Ampère-Maxwell equation ${\bf j}=\nabla\times {\bf B}$ and  the Lorentz force ${\bf j}\times{\bf B}$, which has been used in (\ref{eq_momentum}), are therefore both frame independent.     
On the other hand, the electric field in the rotating frame is obtained by 
Lorentz transforming the electric field in the inertial frame $\boldsymbol{\mathcal{E}}_I$, i.e. 
\begin{equation}
\boldsymbol{\mathcal{E}}=\boldsymbol{\mathcal{E}}_I +({\bf\Omega}\times{\bf r})\times{\bf B}.\label{eq_E_transform}
\end{equation}
Now, the fluid being charge neutral and non-resistive, $\boldsymbol{\mathcal{E}}_I$ is given by the Ohm's law $\boldsymbol{\mathcal{E}}_I= -{\bf u}_I\times{\bf B}_I$, which transforms to 
\begin{equation}
	\boldsymbol{\mathcal{E}}= -{\bf u}\times{\bf B} \label{eq_Ohm}
\end{equation}
in the rotating frame. Consistently, equation (\ref{eq_Ohm}) has been plugged into the steady-state version of Faraday's equation $\nabla\times\boldsymbol{\mathcal{E}}=0$ to obtain (\ref{eq_induction}) in the rotating frame.  Finally, since the electric field is obtained from Ohm's law (\ref{eq_Ohm}), the Maxwell-Poisson equation $\nabla\cdot\boldsymbol{\mathcal{E}}_I=\varrho_I=0$ does not need to be solved. It is however interesting to note that (\ref{eq_E_transform}), implies that its counterpart in the rotating frame is $\nabla\cdot\boldsymbol{\mathcal{E}}=\sigma$, where $\sigma\equiv 2{\bf B}\cdot{\bf \Omega}-({\bf\Omega}\times{\bf r}) \cdot{\bf j}$ 
can be interpreted as fictitious charge density one would have to take into account if the Maxwell-Poisson equation was used to compute the electric field in the rotating frame.  

In the following, we may often want to trace the motion of magnetic field lines. In the frame of ideal MHD, the task is relatively simple, as field lines are frozen in the plasma and are thus displaced perpendicularly to ${\bf B}$ at the velocity   
\begin{equation}
	{\bf u}_{\perp}={\bf u} -\frac{{\bf u}\cdot{\bf B}}{B^2}\,{\bf B}\label{eq_u_perp}   
\end{equation}
which also gives the orientation of the Poynting vector ${\bf S}$, which with (\ref{eq_Ohm}) becomes  
\begin{equation}
	{\bf S}=\boldsymbol{\mathcal{E}}\times {\bf B}={\bf u_\perp} B^2 \label{eq_poynting_vector}.   
\end{equation}

\section{Symmetries\label{sec_symmetries}}

As already graphically illustrated in Figure \ref{fig_vortex_geometry}, 
we restrict the problem to the case where $\bf u$ and $\bf B$ repeat periodically along $z$ and where all azimuthal directions $\varphi$ are equivalent. Accordingly, adopting cylindrical coordinates $(r,\varphi,z)$, we write ${\bf u}={\bf u}(r,\xi)$ and 
${\bf B}={\bf B}(r,\xi)$ where $\xi\equiv \varphi -kz$ and $k$ is the (signed) wave vector specifying the periodicity of the system in the $z$ direction ($|k|=2\pi/\ell$).
In short, the system is invariant under application of the translation  $\xi \rightarrow \xi \pm 2\pi\,n$, with $n \in \mathbb{N}$.

\begin{figure}
	\center{\rotatebox{0}{\resizebox{0.6\textwidth}{!}
			{\includegraphics{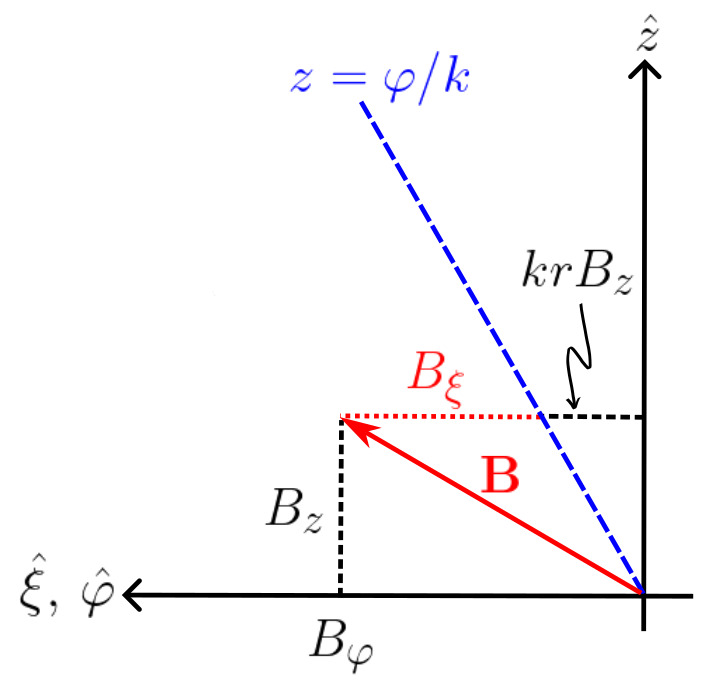}
	}}}
	\caption{Graphical illustration of the components of a position vector $\bf B$ in the $(r,\varphi)$ and $(r,\xi)$ representation. 
	The $\hat{\bf r}$ axis is oriented into the plane of the figure. The curve $z=\varphi/k$ is the direction of invariance of the system.}  
	\label{fig_coordinate_definitions}
\end{figure}

Figure (\ref{fig_coordinate_definitions}) illustrates the geometrical meaning of the coordinate $\xi$ in the $(\varphi,z)$ plane. Whereas $\varphi$ is the angular distance of a point $(r,\varphi,z)$ with respect to a fixed radial axis (for example the $x$ axis), $\xi$ is the angular distance to the $x$ axis after a rotation by 
an angle $kz$ around the $z$ axis. 
Consistently, the components of the vector $\bf B$ can  be defined by using the (generally) non-orthogonal decomposition
\begin{eqnarray}
	{\bf B} =  
	 B_r\hat{\bf r} + B_\xi\hat{\bf\xi} + B_z\begin{pmatrix}0\\kr\\1\end{pmatrix}\label{eq_non_orth_decomp}
\end{eqnarray}
where $\hat{\bf r}$ and $\hat{\bf \xi}$ are unity vectors and $B_\xi\equiv B_\varphi-krB_z$. We shall see, that given the problem's symmetries, the decomposition (\ref{eq_non_orth_decomp}) is better adapted than the standard decomposition using cylindrical coordinate. 
We name the decomposition (\ref{eq_non_orth_decomp}) of a vector field $\bf X$ in the $(r,\xi)$ plane, the $(r,\xi)$ representation 
of $\bf X$.  

\section{Characteristic equations\label{sec_characteristic}}

Partial derivatives with respect to $\varphi$ and $z$ can be expressed in terms of derivatives with respect to  
$\xi$, as $\partial/\partial \varphi=\partial/\partial \xi$ and $\partial/\partial z=-k\partial/\partial \xi$. Recalling that in cylindrical coordinates $\nabla\cdot {\bf u}=r^{-1}\partial_r(ru_r)+r^{-1}\partial_\varphi u_\varphi  +\partial_z u_z=0$, 
equations (\ref{eq_continuity}) and (\ref{eq_maxwell_gauss}) can be written as 
\begin{eqnarray}
	\partial_r(rB_r)  +\partial_\xi B_\xi=0\label{eq_maxwell_gauss_constraint_rxi}\\
	\partial_r(ru_r)+
	\partial_\xi u_\xi =0\label{eq_cont_constraint_rxi}.
\end{eqnarray}
Accordingly, in the $(r,\xi)$ plane, field lines and streamlines are given by the lines of constant value of the 
stream functions $\psi(r,\xi)$ and $\phi(r,\xi)$: 
\begin{eqnarray}
	rB_r&=&\partial \psi/\partial\xi, \; B_\xi=-\partial \psi/\partial r \label{eq_def_psi}\\
	ru_r&=&\partial \phi/\partial\xi, \; u_\xi=-\partial \phi/\partial r.
\label{eq_def_phi}\end{eqnarray}
Defining the 6-dimensional vector 
\begin{equation}
{\bf X}(r,\xi) \equiv \begin{pmatrix}
	B_r(r,\xi)\\
	B_\varphi(r,\xi)\\
	B_z(r,\xi)\\
	u_r(r,\xi)\\
	u_\varphi(r,\xi)\\
	u_z(r,\xi) \end{pmatrix}
\end{equation}	
we can combine equations (\ref{eq_induction}) and (\ref{eq_momentum})
into a system of first order differential equation in the two variables $r$ and $\xi$ which we write in matricial form  
\begin{equation}
	M_r \frac{\partial \bf X}{\partial r}+ M_\xi\frac{\partial \bf X}{\partial \xi} = {\bf A} \label{eq_diff}
\end{equation}
with 
\begin{equation} 
	{\bf A}=\begin{pmatrix}0\\
		{{B_\varphi}} {{u_r}}-{{B_r}} {{u_\varphi}}\\
		0\\
		({{r}} {{\Omega }}+{{u_\varphi}})^2-{{{{B_\varphi}}}^{2}}\\
		-2  \, r \Omega {{u_r}} - \, {{u_r}} {{u_\varphi}}+{{B_r}} {{B_\varphi}}\\
		0\end{pmatrix}\mbox{}\label{eq_A}.
\end{equation}
For completeness, the $6\times6$ square matrices $M_r$ and $M_\xi$ are explicitly given in appendix \ref{app_matrices}. 
If $\det(M_r)={r^6}{u_r}^{2} ({u_r}-{B_r})^2 ({u_r}+{B_r})^2\neq 0$, one may cast (\ref{eq_diff}) into the normal form 

\begin{equation}
	\frac{\partial \bf X}{\partial r}+ \tilde{M}_\xi\frac{\partial \bf X}{\partial \xi} = \tilde{\bf A}
	\label{eq_diff_normal}
\end{equation}
where $\tilde{M}_\xi\equiv M_r^{-1}M_\xi$ 
and 
\begin{equation} 
	\tilde{\bf A}\equiv M_r^{-1}{\bf A}=\frac{1}{r u_r (u_r^2 -B_r^2)}\begin{pmatrix}
		 [-B_\varphi(B_r A_5+u_r A_2) + A_4(u_r^2 -B_r^2)]B_r/u_r\\
		 u_r(B_r A_5+u_r A_2)\\
		 0\\
		-B_\varphi(B_r A_5+u_r A_2) + A_4(u_r^2 -B_r^2)\\
		 u_r(u_r A_5+B_r A_2)\\
		0\end{pmatrix}\mbox{}.\label{eq_A_tilde}
\end{equation}

For an eigenvalue $\lambda$ and associated eigenvector ${\bf E}$ of $\tilde{M}$, we can formally write $\tilde{M}_\xi{\bf E} =\lambda {\bf E}$ and the corresponding characteristic equations 
\begin{eqnarray}
	\frac{\partial \bf X}{\partial r} &=& \tilde{\bf A}-\lambda{\bf E}\label{eq_caract_r}\\
	 \frac{\partial \bf X}{\partial \xi}&=&{\bf E} \label{eq_caract_xi} 
\end{eqnarray}
from where it follows that the component $X_i$ is constant along the characteristic trajectory defined by  
\begin{equation}
	\left(\frac{d\xi}{d r}\right)_i = \lambda -\frac{\tilde{A}_i}{E_i}
	\label{eq_dxi_dr_i}
\end{equation}	
a property one may use to experimentally determine $\lambda$ and $\tilde{A}_i$. For example, 
since $\tilde{A}_3=0$, $\lambda$ can be obtained by measuring the orientation of the contours of constant $B_z$ in the $(r,\xi)$ plane. 

Equations (\ref{eq_caract_r}) and (\ref{eq_caract_xi}) have been obtained without the constraints imposed by both the Maxwell-Gauss equation (\ref{eq_maxwell_gauss_constraint_rxi}) and the continuity equation (\ref{eq_cont_constraint_rxi}). The two equations effectively imply the following relations:   
\begin{eqnarray}
	B_r=-r(\tilde{A}_1-\lambda E_1) - (E_2 -kr E_3)
	\label{eq_maxwell_gauss_constraint}\\
	u_r=-r(\tilde{A}_4-\lambda E_4) - (E_5 -kr E_6).\label{eq_cont_constraint}
\end{eqnarray}

\section{Eigenvalues and eigenvectors\label{sec_eigen}}
  
The eigenvalues and associated eigenvectors of $\tilde{M}_\xi$ fall  in two classes.

\begin{equation}
\lambda_0= \frac{u_\varphi-kru_z}{ru_r},\; {\bf E}_0=\begin{pmatrix} u_\varphi-kr\,
		u_z\cr -u_r\cr kr\,u_r\cr 0\cr 0\cr 0\cr 
\end{pmatrix} \label{eq_eigen_0}
\end{equation} 
and   
\begin{equation}
\lambda = \frac{(Pu_\varphi-krPu_z)-(B_\varphi -krB_z)}{r(Pu_r-B_r)},\; {\bf E}_{a}=\begin{pmatrix} 
	    B_z\cr 0\cr -B_r\cr 
	    PB_z\cr 0\cr -PB_r\cr 
		\end{pmatrix},\; 
{\bf E}_{b}=\begin{pmatrix} 
	       0 \cr B_z\cr -B_\varphi\cr 
	       0 \cr PB_z\cr -PB_\varphi\cr 
		\end{pmatrix} \label{eq_eigen_1}
\end{equation}
where $P=\pm 1$ correspond to two Alfvénic modes for which the magnetic and velocity fluctuations vary in phase ($P=1$) or in antiphase ($P=-1$), respectively.

\section{Structure of the axisymmetric eigenmode\label{sec_axisymmetric_mode}}

For the eigenmode (\ref{eq_eigen_0}) one has $E_4=E_5=E_6=0$, i.e. $\partial_\xi \bf u=0$ so that from (\ref{eq_cont_constraint_rxi}) and the constraint of a finite value $u_r(r=0)$, it follows that $u_r=0$ everywhere. From (\ref{eq_eigen_0}) one  also deduces $E_2=E_3=0$, which then also implies $\partial_r(rB_r)=0$ and, therefore, $B_r=0$ everywhere. As a consequence, $\partial_\xi {\bf X}={\bf E}=0$, and the mode does not depend on $\xi$ which excludes the possibility of a singularity other than at $r=0$. 
Finally, given $B_r=u_r=0$, only the fourth row of the original system of equations (\ref{eq_diff}) is non-zero and reads 
\begin{eqnarray}
	B_\varphi\partial_r B_\varphi + B_z\partial_r B_z 
	&=& [(r\Omega+u_\varphi)^2 -B_\varphi^2]/r\nonumber \\ 
	&=& r\Omega^2 +2\Omega u_\varphi+u_\varphi^2/r -B_\varphi^2/r 
	\label{eq_lambda_0_equilibrium}
\end{eqnarray}
with all quantities depending solely on $r$. On the right, in equation (\ref{eq_lambda_0_equilibrium}) one recognizes the magnetic pressure gradient term on the left, and the centrifugal, Coriolis, inertia and magnetic tension terms, respectively. We shall not consider this mode any farther. 

\section{Structure of the Alfvénic  eigenmode\label{sec_alfven_mode}}

A general property of the eigenvectors (\ref{eq_eigen_1}) is that $\sum_{i=1}^3 E_i X_i=0$, from where one  concludes that the magnetic pressure dos not depend on $\xi$:     
\begin{equation}
	\frac{\partial }{\partial \xi}\left(\frac{B^2}{2}\right)= 
	 \sum_{i=1}^3 E_i X_i=0\label{eq_d_dxi_B}.
\end{equation}
From (\ref{eq_caract_r}) it then follows that the radial variation of the magnetic pressure is given by 
\begin{equation}
	\frac{\partial }{\partial r}\left(\frac{B^2}{2}\right)= 
	B_r \tilde{A}_1+B_\varphi\tilde{A}_2\label{eq_d_dr_B}  
\end{equation}
and, similarly, for the kinetic energy 
\begin{equation}
	\frac{\partial }{\partial r}\left(\frac{u^2}{2}\right)= 
	u_r \tilde{A}_4+u_\varphi\tilde{A}_5
	-\lambda\sum_{i=4}^{6} E_i X_i.\label{eq_d_dr_u}  
\end{equation}
Hereafter we specifically investigate the general structure of the eigenmode around a vortex type singularity.

\subsection{Postulating the existence of a singularity}

We postulate the existence of a singular, static (in the rotating frame) magnetic field line located at  $(r_0,\xi_0)=(1,0)$. 
From the field line equation $\varphi-kz=\xi_0$ and $r=r_0$ we deduce that $B_{\varphi0}/B_{z0}=r_0d\varphi/dz=kr_0$. 
In addition, for the field line not to be displaced by the flow,  ${\bf u}_0$ must be parallel to ${\bf B}_0$, which implies $u_{r0}=0$ and $u_{\varphi0}=kr_0 u_{z0}$. In summary, the singularity is defined by the following conditions: 
\begin{equation}\begin{matrix}
		B_{r0}&=& 0\\
		B_{\xi0}&=& B_{\varphi 0}-kr_0\,B_{z0} = 0 \\
		u_{r0}&=& 0\\
		u_{\xi0}&=& u_{\varphi 0}-kr_0\,u_{z0} = 0 \\
	\end{matrix}\label{eq_X0_singularity}.
\end{equation}	  
We note that with these assumptions and (\ref{eq_A}), at the singularity, the only possibly non-zero term of the vector $\bf A$ is the component $A_4$:  
\begin{equation}
	r_0 \partial_r(\tfrac{1}{2}B_{\varphi}^2 + \tfrac{1}{2}B_{z}^2)=A_4= 
	({{r_0}} {{\Omega }}+{{u_{\varphi 0}}})^2-{{{{B_{\varphi 0}}}}^{2}}.\label{eq_A_40}
\end{equation} 
One may be tempted to set $A_4=0$, thus assuming an extremum of the magnetic pressure in the radial direction and, by doing so, forcing a relation between $u_{\varphi 0}$ and $B_{\varphi 0}$. The condition $A_4=0$ also corresponds to an equilibrium between  inertia, magnetic tension, Coriolis force and centrifugal force.    
We made such an assumption in a previous paper by  \cite{Pantellini_2020}, but in general, one has to admit that $A_4\neq 0$.  
We conclude this section by noting that at the singularity, ${\rm det}M_r=0$ so that the normal form (\ref{eq_diff_normal}) is undefined there. Accordingly, except for $\tilde{A}_3$ and $\tilde{A}_6$ which are both zero, all components of $\tilde{\bf A}$ and the eigenvalues of the system reduce to an indeterminate $0/0$ ratio which we assume to be finite for the singularity to be a regular one. 

\subsection{First order structure near the singularity}

Given that both eigenvectors ${\bf E}_a$ and ${\bf E}_b$ are associated with the same eigenvalue $\lambda$, any linear combination of ${\bf E}_a$ and ${\bf E}_b$ is also an eigenvector for the eigenvalue $\lambda$ \citep[e.g.][]{Hefferon_2017}. To the lowest order, the most general form of the eigenvector near the singularity can then be written as  
\begin{equation}
	{\bf E} = a{\bf E}_a + b{\bf E}_b \simeq 
	\begin{pmatrix} 
		aB_{z 0}\cr bB_{z 0}\cr -bB_{\varphi 0} \cr 
		aPB_{z 0}\cr bPB_{z 0}\cr -bPB_{\varphi 0}\cr 
	\end{pmatrix}\label{eq_E}
\end{equation} 
where $a$ and $b$ are constants which, according to 
(\ref{eq_caract_xi}), define (or are defined by) the variations of the magnetic field components with respect to $\xi$ at the singularity: 
\begin{eqnarray}
	a &=& B_{z0}^{-1}\frac{\partial  B_r}{\partial \xi} = 
	      P\,B_{z0}^{-1}\frac{\partial  u_r}{\partial \xi}\label{eq_par_a}\\ 
	b &=& B_{z0}^{-1}\frac{\partial  B_\varphi}{\partial \xi} =
	      P\,B_{z0}^{-1}\frac{\partial  u_\varphi}{\partial \xi} \nonumber \\  &=&
	      \frac{B_{z0}^{-1}}{1+(kr_0)^2}\frac{\partial  B_\xi}{\partial \xi}.\label{eq_par_b} 
\end{eqnarray} 
Expanding ${\bf X}={\bf X}_0+\delta{\bf X}$, where  
${\bf X}_0$ are the values at the singularity and $\delta{\bf X}$ the first order corrections in $\delta r$ and $\xi$, we can write
\begin{eqnarray}
	\frac{\partial }{\partial r}\, \delta{\bf X} &=& \tilde{\bf A}_0-\lambda{\bf E}\label{eq_caract_first_r}\\
	\frac{\partial }{\partial \xi}\,\delta{\bf X}&=&{\bf E}. \label{eq_caract_first_xi} 
\end{eqnarray}
Using either (\ref{eq_maxwell_gauss_constraint}) or (\ref{eq_cont_constraint}), the eigenvalue $\lambda$ can then be evaluated explicitly at the singularity: 
\begin{equation}
	\lambda = \frac{\tilde{A}_{10}}{a B_{z0}} +
	r_0^{-1}\,\frac{b}{a}
	[1+(r_0k)^2]\label{eq_lamda_alpha}.
\end{equation}
Formally, the first order solution at the singularity is then given by 
\begin{equation}
	{\bf X} = {\bf X}_0 +({\bf A}_0-\lambda{\bf E})\delta r +{\bf E}\xi
	\label{eq_X_first_order_solution}.
\end{equation}
where only the non-zero components of ${\bf A}_0$ remain  to be determined explicitly given that their expressions  in (\ref{eq_A_tilde}) are of the type $0/0$ at the singularity.   

\subsection{Evaluating $\tilde{\bf A}_0$ at the singularity\label{sec_A}}

From (\ref{eq_A_tilde}) we do already know that $\tilde{A}_{30}=\tilde{A}_{60}=0$. 
From (\ref{eq_d_dr_B}) and (\ref{eq_A_40}) we also 
know that for finite $B_{\varphi 0}$:     
\begin{equation}
	\tilde{A}_{20} = A_4/B_{\varphi 0} =[(r_0 \Omega +u_{\varphi 0})^2-B_{\varphi 0}^2]/B_{\varphi 0}.
	\label{eq_A20_tilde}
\end{equation}
From (\ref{eq_E}) we deduce that, at the singularity, all eigenmodes satisfy $\partial_\xi \delta{\bf B} = P\partial_\xi \delta{\bf u}$ and thus, from the zero divergence conditions (\ref{eq_cont_constraint_rxi}) and (\ref{eq_maxwell_gauss_constraint_rxi}), that $\partial_r  B_r=P \partial_r u_r$, so that  
\begin{equation}
	B_r(r,\xi)=Pu_r(r,\xi) \;\text{up to first order in}\;\delta r=r-r_0 \text{ and } \xi. \label{eq_Br_Pur}
\end{equation}
We also note that equations  (\ref{eq_maxwell_gauss_constraint}) and (\ref{eq_cont_constraint}) imply that 
\begin{eqnarray}
	\tilde{A}_{40}&=&P\tilde{A}_{10}
\end{eqnarray}
with no a priori constraint on $\tilde{A}_{10}$ which we  therefore consider to be a free parameter. 

Given the structure of the eigenvector (\ref{eq_E}), we can evaluate the radial variations of both the magnetic pressure and the kinetic energy at the singularity:   
\begin{eqnarray}
	\frac{\partial}{\partial r} \left(\frac{B^2}{2}\right)&=& B_{\varphi 0} \tilde{A}_{20}\label{eq_dr_B2}\\  
	\frac{\partial}{\partial r} \left(\frac{u^2}{2}\right) &=&
	u_{\varphi 0} \label{eq_dr_u2}\tilde{A}_{50}  
\end{eqnarray} 
where $\tilde{A}_{i0}\equiv \tilde{A}_i(r_0,0)$. 
From (\ref{eq_A_tilde}) we deduce that 
\begin{equation}
	\tilde{A}_{50}=P\tilde{A}_{20}-\Omega\label{eq_A50_A20}
\end{equation}
which finally implies that the velocity and magnetic fluctuations are related through  
\begin{equation}
	\delta{\bf B}=P(\delta{\bf u}+\Omega\delta r\,\hat{ \varphi})\label{eq_dB_du}
\end{equation}
where $\hat{\varphi}$ is the unity vector in the direction of the $\varphi$ coordinate. Equation (\ref{eq_dB_du}) completes the already established relation (\ref{eq_Br_Pur}) for the components $\delta B_r$ and $\delta u_r$.

\subsection{Parameters defining the Alfvén vortex}

Without specification of the conditions elsewhere than at the singularity, the structure of the eigenmode in the singularity's vicinity is fully specified through the following parameters: 
\begin{equation}
	\Omega,\; k,\;B_{\varphi 0},\; u_{\varphi 0},\;
	a,\; b,\;\tilde{A}_{10},\;P=\pm 1 \label{eq_problem_parameters}. 
\end{equation}
Hence, within the frame of ideal and incompressible MHD, the linear structure of a static Alfvén vortex is  described by the  parameters listed in (\ref{eq_problem_parameters}).
Inspired by (\ref{eq_A20_tilde}), we may define the dimensionless parameter $\Lambda\equiv 1+\tilde{A}_{20}/B_{\varphi 0}$ to further characterizes the vortex  
\begin{equation}
	\Lambda= 
	\frac{u_{\varphi 0}^2}{B_{\varphi 0}^2}
	\left(
	\frac{r_0\Omega}{u_{\varphi 0}}+1
	\right)^2 
\;
\begin{cases}=0 \,:\; r_0\Omega+u_{\varphi 0}=0\\
	<1\,:\;\partial_r B^2 < 0 \\
	>1\,:\;\partial_r B^2 > 0
\end{cases}	
	.\label{eq_Lambda}
\end{equation}  
wherein one recognizes the Rossby number, which we conveniently define as ${\rm Ro}\equiv|u_{\varphi 0}/(r_0\Omega)|$. For ${\rm Ro}\gg 1$ the Coriolis force acting on the fluid dominates over the centrifugal force at the vortex center. The opposite is true for ${\rm Ro}\ll 1$. Note that in (\ref{eq_Lambda}), the limiting case $\Lambda=0$ corresponds to the case where the fluid at the vortex center is exactly corotating.

\subsection{Stream functions near the singularity}

Given the set of parameters (\ref{eq_problem_parameters}), the stream functions $\psi$ and $\phi$ defined in (\ref{eq_def_psi}) and (\ref{eq_def_phi}) can now be computed explicitly in the vicinity of the singularity up to second order in $\xi$ and $\delta r=r-1$. For the stream function associated to $\bf B$ we thus write   
\begin{eqnarray}
	\psi(r,\xi)= \tfrac{1}{2}\psi_{\xi\xi}\,\xi^2 + \psi_{r \xi}\,\delta r\xi +\tfrac{1}{2}\psi_{rr}\,\delta r^2 
	\label{eq_psi_stream_function}
\end{eqnarray}
which is the equation of a conic section with main axis rotated by $\theta$ with respect to the $r$ axis:
\begin{eqnarray}
	\theta = \frac{1}{2}\arctan\left( \frac{2\,\psi_{r\xi}}{\psi_{rr}-\psi_{\xi\xi}}\right)+
	  \begin{cases}
		0 \;    &\text{for}\;\psi_{rr}>\psi_{\xi\xi} \\
		\pi/2\; &\text{for}\; \psi_{rr}<\psi_{\xi\xi} 
	\end{cases}	
	\label{eq_theta} 
\end{eqnarray}   
Using  (\ref{eq_def_psi}), (\ref{eq_caract_r}) and (\ref{eq_caract_xi}), we easily find the (constant) coefficients to be       
\begin{eqnarray}
\psi_{\xi\xi} &=& E_1 = a B_{z 0} \nonumber \\
\psi_{r\xi}   &=& \tilde{A}_1 - \lambda E_1 
               = -b B_{z 0} (1+k^2) 
\label{eq_coeff_psi}\\
\psi_{rr}     &=& k(B_{z0}-\lambda E_3)
-(\tilde{A}_2-\lambda E_2) \nonumber \\ 
&=& kB_{z 0}-\tilde{A}_2 +\lambda b B_{z 0} (1+k^2)  \nonumber
\end{eqnarray} 
where all quantities are obviously evaluated at $(r,\xi)=(r_0,0)$. 
The topology of the field lines in the $(r,\xi)$ plane depend on the value of the discriminant  
\begin{eqnarray}
 D_B &\equiv& \psi_{r \xi}^2 -	\psi_{\xi\xi}\psi_{rr}\nonumber  \\ &=& 
   \frac{a}{k}\left[(\Omega+u_{\varphi 0})^2-2B_{\varphi 0}^2\right]-\tilde{A}_{1} bB_{z0}(1+k^2)\;
  \begin{cases}<0\text{ ellipse}\\
  	           =0\text{ parabola}\\
  	           >0\text{ hyperbola}
  \end{cases}\label{eq_disc_B}
\end{eqnarray} 
In case of an ellipse, the ratio of its semi-axes can also be expressed in terms of the coefficients of the stream function $\psi$:  
\begin{equation}
	\frac{\mathcal{L}_1^2}{\mathcal{L}_2^2}=
	\frac{\psi_{\xi\xi}-2\psi_{r\xi}\tan\theta+\psi_{rr}\tan^2\theta}
	     {\psi_{rr}+2\psi_{r\xi}\tan\theta+\psi_{\xi\xi}\tan^2\theta}
	     \label{eq_P1_P2}
\end{equation}
where $\theta$ is given by (\ref{eq_theta}).
Similarly, the stream function $\phi$ must be of the type 
\begin{equation}
	\phi(r,\xi)= \tfrac{1}{2}\phi_{\xi\xi}\,\xi^2 + \phi_{r \xi}\,\delta r\xi +\tfrac{1}{2}\phi_{rr}\,\delta r^2 
\label{eq_phi_stream_function}
\end{equation}
with 
\begin{eqnarray}
	\phi_{\xi\xi} &=& P\,\psi_{\xi\xi} \nonumber \\
	\phi_{r\xi}   &=& P\,\psi_{r\xi} \label{eq_coeff_phi} \\
	\phi_{rr}     &=& P\,\psi_{rr}-(P\,B_{\varphi 0}-u_{\varphi 0}-\Omega)\nonumber  
\end{eqnarray}
As for the magnetic field, the kind of curve described by the motion of a fluid parcel in the $(r,\xi)$ plane depends on the value of the corresponding discriminant:   
\begin{eqnarray}
	D_u &\equiv& \phi_{r \xi}^2 -	\phi_{\xi\xi}\phi_{rr}\nonumber \\
	&=&D_B+\frac{a PB_{\varphi 0}}{k}(PB_{\varphi 0}-u_{\varphi 0}-\Omega)\;
	\begin{cases}<0\text{ ellipse}\\
		=0\text{ parabola}\\
		>0\text{ hyperbola}
	\end{cases}\label{eq_disc_u}
\end{eqnarray}  
from where it appears that the topology of magnetic field lines can be different from the topology of the flow lines as $D_B$ and $D_u$ may have different signs.  
The displacement velocity  ${\bf u}_\perp$ of a field line (see (\ref{eq_u_perp})) is also of some interest because of its connection with the Poynting vector (see (\ref{eq_poynting_vector})). Since ${\bf u}_\perp$ is generally not divergence-free, one may not be allowed to use a stream function as for $\bf u$ and $\bf B$. However, to first order one can write (see (\ref{eq_u_perp_approx}) below) 
\begin{eqnarray}
	u_{\perp,r} &=& u_r - \frac{u_{\varphi 0}}{B_{\varphi 0}}\, B_r\\ 
	u_{\perp,\xi} &=& u_\xi - \frac{u_{\varphi 0}}{B_{\varphi 0}}\, B_\xi 
\end{eqnarray}
so that for ${\bf u}_\perp$ we may conveniently use the stream function 
\begin{equation}
	 \zeta = \phi-\frac{u_{\varphi 0}}{B_{\varphi 0}} \psi \label{eq_zeta_stream_function}
\end{equation}
with the associated discriminant $D_{u\perp}\equiv \zeta_{r \xi}^2 -	\zeta_{\xi\xi}\zeta_{rr}$ becoming   
\begin{eqnarray}
	D_{u\perp} =D_B\frac{(P\,B_{\varphi 0}-u_{\varphi 0})^2}{B_{\varphi 0}^2}
	- \frac{a}{k}(P\,B_{\varphi 0}-u_{\varphi 0}) 
	(P\,B_{\varphi 0}-u_{\varphi 0}-\Omega)
	\;
	\begin{cases}<0\text{ ellipse}\\
		=0\text{ parabola}\\
		>0\text{ hyperbola}
	\end{cases}\label{eq_disc_uperp}
\end{eqnarray}  

\subsection{An example of linear vortex}

We define a vortex by the set of parameters in Table \ref{tab_vortex_param}.
\begin{table}[ht]
		\begin{tabular}{c|c|c|c|c|c|c|c}
			$r_0\Omega$ & $r_0k$ & $B_{\varphi 0}$ & $u_{\varphi 0}$ & $a$ &  $b$ & $r_0\tilde{A}_{10}$ & $P$ \\
			\hline\rule[0ex]{0pt}{2.5ex}
			$-1/4$& $1$ & $1/2$ & $1$ & $2$ & $1/2$ & $2$ & $+1$\\
		\end{tabular}
		\caption{Vortex parameters (normalized units) used for the Figures \ref{fig_r_xi} and \ref{fig_example_field_and_stream_lines}.}\label{tab_vortex_param}
\end{table}

The corresponding derived parameters (some of which will be defined later) are given in Table \ref{tab_vortex_structure}. 

\begin{table}[ht]
		\begin{tabular}{c|c|c|c|c|c|c|c|c}
			& $D_i$ &$\theta$ [deg] & $\mathcal{L}_1/\mathcal{L}_2$ & $s_1$ & $r_0\lambda$ &$\Lambda$& Ro & $v_{{\rm ph},z}$\\
			\hline\rule[0ex]{0pt}{2.5ex}
			$B$ & $-7/8$ & $48.6$& $1.67$ &  $4.75$ & $5/2$ & $9/4$ & $4$ & $4$\\
			\hline\rule[0ex]{0pt}{2.5ex}
			$u$ & $-9/8$ & $55.3$ & $1.62$ &  $8.38$ & $''$ & $''$ & $''$ & $''$\\
			\hline\rule[0ex]{0pt}{2.5ex}
			$u_{\perp}$ & $-5/8$ &$131.4$& $0.55$ &  $-$ & $''$ & $''$ & $''$& $''$
			
		\end{tabular}
		\caption{Derived parameters for the vortex defined in Table \ref{tab_vortex_param} pertinent to Figures \ref{fig_r_xi} and \ref{fig_example_field_and_stream_lines}. $s_1$ is the distance along the streamline for the streamline to complete one turn  around the vortex center, and $v_{{\rm ph},z}$ is the phase speed of the vortex in the non-rotating frame (see (\ref{eq_v_phz})).}
		\label{tab_vortex_structure}
\end{table}
From the table it appears that the discriminants $D_B$, $D_u$ and $D_{u\perp}$ are all negative, implying that streamlines are elliptical for all fields. 
\begin{figure}
	\center{\rotatebox{0}{\resizebox{0.6\textwidth}{!}
			{\includegraphics{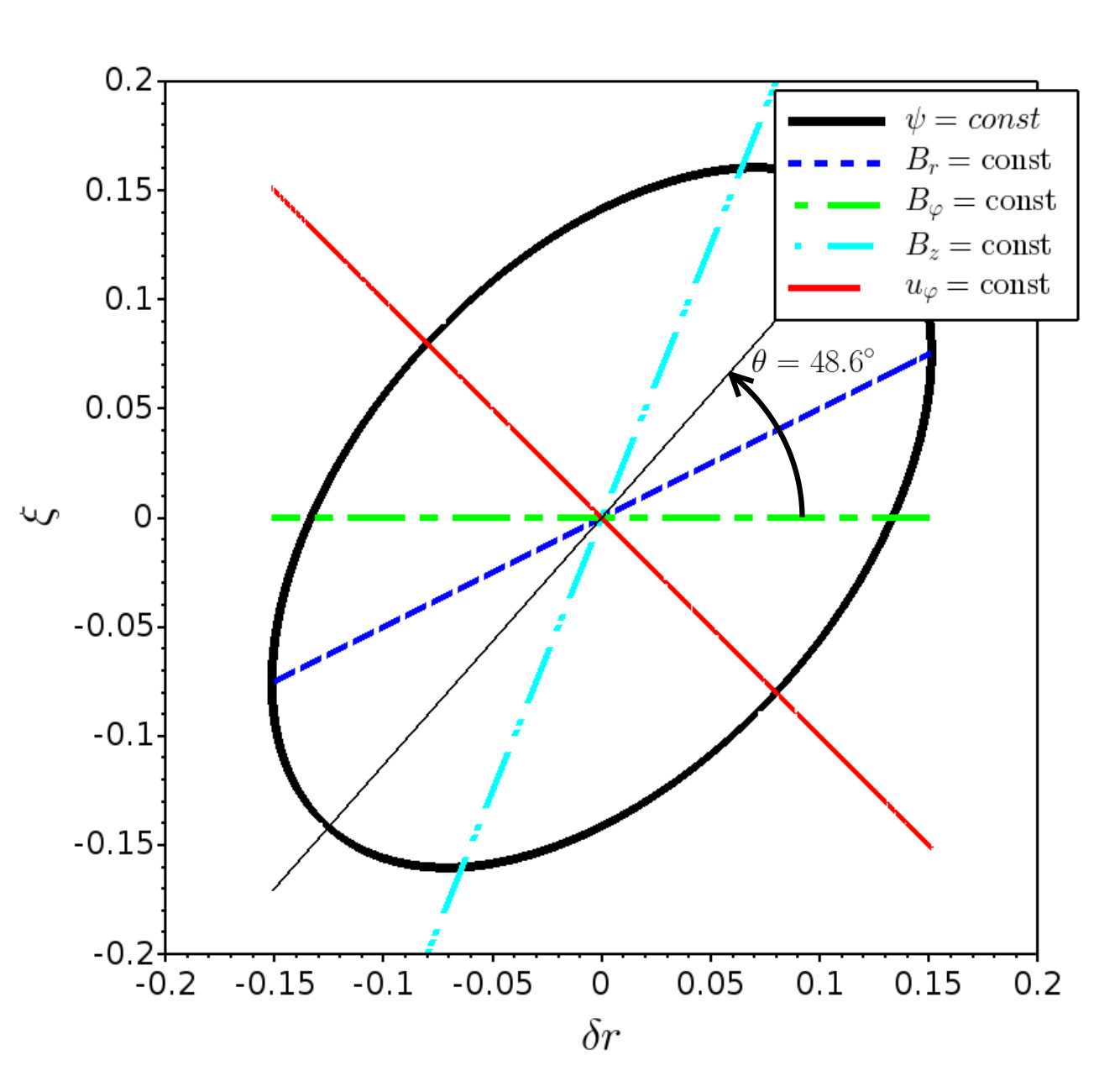}
	}}}\caption{Magnetic field stream function $\psi(\delta r,\xi)$ in the vicinity of the vortex defined in Table {\ref{tab_vortex_param}}. 
Shown are also the directions of invariance for some selected field components.}
	\label{fig_B_stream_and invariant_dirs}
\end{figure}
As an example, a magnetic streamline is shown in Figure \ref{fig_B_stream_and invariant_dirs}. Following  (\ref{eq_theta}), the ellipse is rotated anti-clockwise by an angle $\theta=48.6^\circ$ with respect to the $r$ axis. The directions of invariance given by (\ref{eq_dxi_dr_i})
are also plotted for some selected field components. 
Directions of invariance can be useful to infer the vortex parameters from experimental data, for example in numerical simulations. 

\begin{figure}
	\center{\rotatebox{0}{\resizebox{0.9\textwidth}{!}
			{\includegraphics{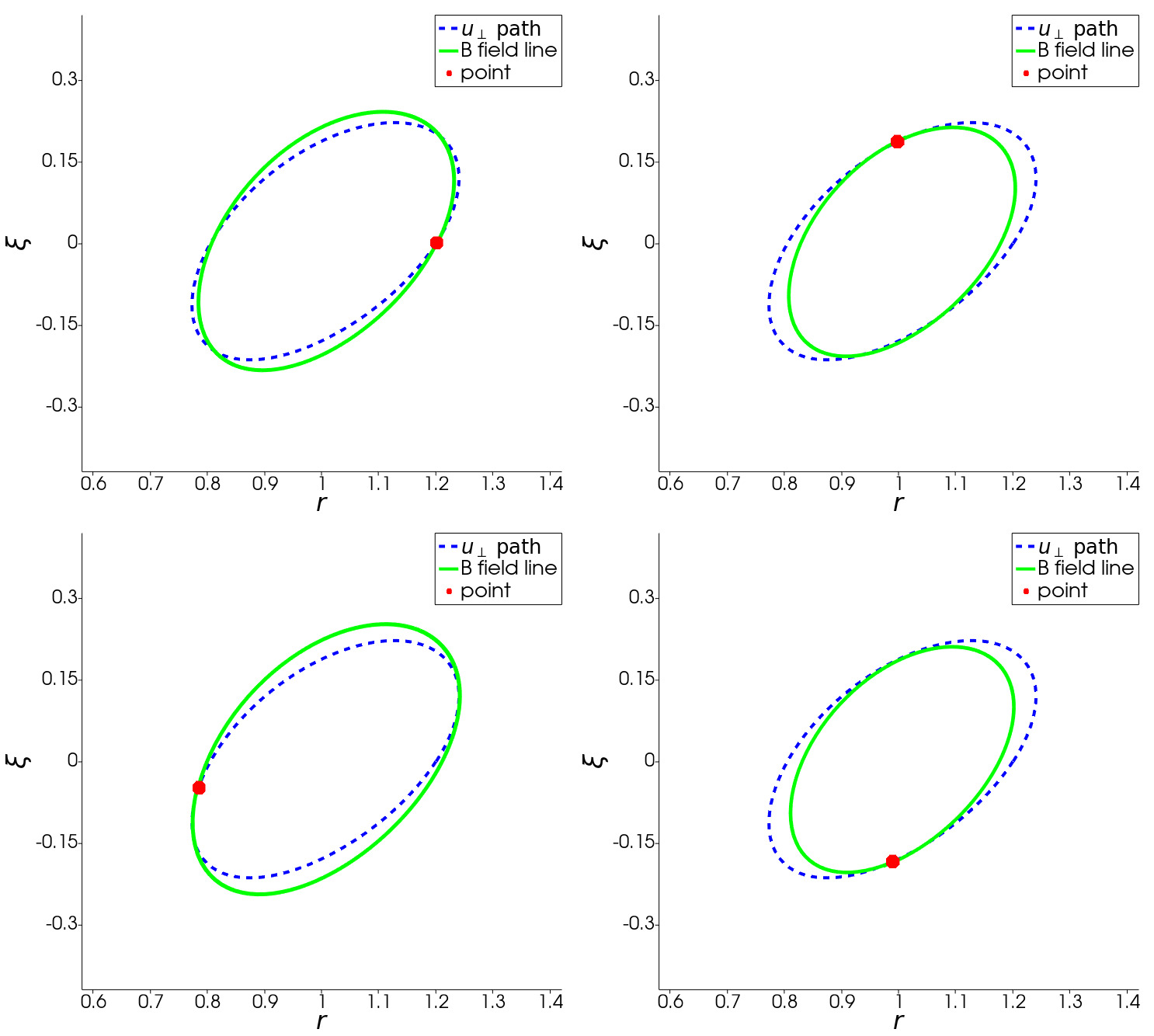}
	}}}
	\caption{Time evolution	
		 of a magnetic field line (streamline in green) in the $(r,\xi)$ plane representation. The dashed line corresponds to the trajectory of a point on the field line as specified by the stream function of ${\bf u}_\perp$ (\ref{eq_zeta_stream_function}).} 
	\label{fig_r_xi}
\end{figure}

Figure \ref{fig_r_xi} shows how the $(r,\xi)$ representation of a magnetic streamline varies in time as the field line moves around the vortex. The dashed ellipse shows the path described by a point on the magnetic field line, i.e. the streamline for ${\bf u}_\perp$. Since the magnetic streamline must cross the same point while conserving its geometry given by (\ref{eq_psi_stream_function}), it is subject to a periodic homothetic transformation. 

\subsection{3D structure of the magnetic filed lines in the vicinity of the vortex \label{sec_3D_field_lines}}
The stream functions are an excellent means to characterize the nature of a vortex for a given set of parameters in the $(r,\xi)$ plane. To recover the 3D structure of field lines or streamlines, a system of differential equations has to be integrated. For the magnetic field the system is   
\begin{equation}
	\frac{dr}{ds} = \frac{B_r}{B},\; 
	\frac{d\varphi}{ds} = \frac{B_\varphi}{rB},\; 
	\frac{dz}{ds} = \frac{B_z}{B} \label{eq_field_lines}
\end{equation}
where $s$ is the distance along the field line, $B=d{\bf s}\cdot{\bf B}/\lVert d{\bf s}\rVert$, and where ${\bf B}(r,\xi)$ is known through its first order approximation (see (\ref{eq_X_first_order_solution})):  
\begin{equation}
\begin{pmatrix} 
	B_r \cr B_\varphi\cr B_z 
\end{pmatrix}
=
\begin{pmatrix} 
	0 \cr B_{\varphi 0}\cr B_{z 0} 
\end{pmatrix}
+\left[
\begin{pmatrix} 
	A_{10} \cr A_{20}\cr 0 
\end{pmatrix}
-\lambda 
\begin{pmatrix} 
	E_{1} \cr E_{2}\cr E_{3} 
\end{pmatrix}
\right]\delta r
+
\begin{pmatrix} 
	E_{1} \cr E_{2}\cr E_{3} 
\end{pmatrix}\xi. \label{eq_B_first_order}
\end{equation}
The integration is best done numerically, but it is instructive to carry out the integration analytically in order to obtain an expression for the length one has to travel along a field line to complete a turn around the vortex center.  
Assuming, the field line is elliptical in the $(r,\xi)$ plane (otherwise there is no winding), it is convenient to use its parametric description 
\begin{eqnarray}
	\delta r(\alpha)&=& \mathcal{L}_1\cos\theta\cos\alpha
	                  -\mathcal{L}_2\sin\theta\sin\alpha
	                  \label{eq_dr_alpha}\\
	\xi(\alpha)     &=&  \mathcal{L}_1\sin\theta\cos\alpha
	                  +\mathcal{L}_2\cos\theta\sin\alpha
	                  \label{eq_dxi_alpha}
\end{eqnarray} 
where $\alpha$ is the angle with respect to the radial direction while $\mathcal{L}_1$ and $\mathcal{L}_2$ are the length of the ellipse's two half-axis. For $\delta r,\xi \rightarrow 0$ the incremental distance $ds$ along the vortex axis (i.e. along ${\bf B}_0$) is given by 
\begin{equation}
	ds=dr\frac{B}{B_r}\simeq 
	 \frac{\partial\delta r}{\partial\alpha} \frac{B_0\, d\alpha}
	 {(\tilde{A}_{10}-\lambda E_1)\delta r +E_1\xi}
\end{equation}
which can be integrated analytically to obtain $s(\alpha)$. 
The general solution is relatively involved and of little interest. It is however instructive to compute the distance $s_1$ required to complete one turn around the vortex (with $B_0=B_{z0}(1+k^2)^{1/2}$)
\begin{equation}
	s_1=s(2\pi) \simeq  
	\frac{-2\pi\sqrt{1+k^2}\,a\,\tfrac{\mathcal{L}_1}{\mathcal{L}_2}}
	{(\tfrac{\mathcal{L}_1}{\mathcal{L}_2})^2 
	\left[a\sin\theta -b(1+k^2)\cos\theta\right]^2 + 
	\left[a\cos\theta +b(1+k^2)\sin\theta\right]^2}\label{eq_s1}   
\end{equation}
where $\theta$ is given by (\ref{eq_theta}) and ${\mathcal{L}_1}/{\mathcal{L}_2}$ by (\ref{eq_P1_P2}).
As an example, for  circular polarization $\mathcal{L}_1/\mathcal{L}_2=1$ and $b=0$ one has 
\begin{equation}
	|s_1|=\frac{2\pi\sqrt{1+k^2}}{\sqrt{ak}}
\end{equation}
which shows that the number of windings over a given distance along the vortex increases as $\sqrt{a}$. 
Expression (\ref{eq_s1}) also applies to the streamline of $\bf u$ by 
replacing the pertinent values for $\theta$ and ${\mathcal{L}_1}/{\mathcal{L}_2}$ and by multiplication of the expression by the factor $u_0/B_0$. As a consequence, the winding of the magnetic field lines and the plasma streamlines can be very different. In the case of the vortex defined in Table \ref{tab_vortex_param}, the magnetic field lines are nearly two times more twisted than the fluid streamlines as $s_1=4.75$ for $\bf B$ and  $s_1=8.38$ for $\bf u$ (see Table \ref{tab_vortex_structure}). Some authors of the solar physics community \citep[e.g.][]{Silva_etal_2021} 
classify such a magnetic twist dominated vortex as a $M$-vortex and use the denomination $k$-vortex for the vortices dominated by the twist of the kinetic flow.  

\subsection{Trajectory of a fluid parcel and Poynting vector \label{sec_3D_traj}}
The procedure used to compute the magnetic field lines  can obviously be used to compute the 3D structure of  
the streamlines for $\bf u$ and $\bf u_\perp$. 
For the streamlines of ${\bf u}_\perp$ one may use 
its definition (\ref{eq_u_perp}) or its first order approximation     
\begin{equation}
	{\bf u}_\perp\simeq
	\left(
	 \delta{\bf u}-\frac{u_{\varphi 0}}
	 {B_{\varphi 0}}\;\delta{\bf B}
	\right)-
	\left[	\left(
	\delta{\bf u}-\frac{u_{\varphi 0}}{B_{\varphi 0}}\;\delta{\bf B}
    \right)\cdot\frac{{\bf B}_0}{B_0}\right]\frac{{\bf B}_0}{B_0}
	\label{eq_u_perp_approx}
\end{equation}	
where $\delta{\bf u}$ and $\delta{\bf B}$ are tied by the  relation (\ref{eq_dB_du}) and where we have used 
${\bf u}_0/u_{\varphi 0}={\bf B}_0/B_{\varphi 0}$.
Accordingly, to first order, the Poynting vector is 
\begin{equation}
	{\bf S} \simeq B_0^2 \,{\bf u}_{\perp} \label{eq_S_order_1} 
\end{equation}
which implies that there is no first order Poynting vector component parallel to the vortex axis. However, as for the shear-Alfvén wave, a non-zero second order component exists:   
\begin{equation}
	S_{\parallel}={\bf S}\cdot {\bf B}/\vert B\vert \simeq \vert B_0\vert\, {\bf u}_\perp\cdot\delta {\bf B}. \label{eq_S_par}  
\end{equation}

\begin{figure}
	\center{\rotatebox{0}{\resizebox{0.8\textwidth}{!}
			{\includegraphics{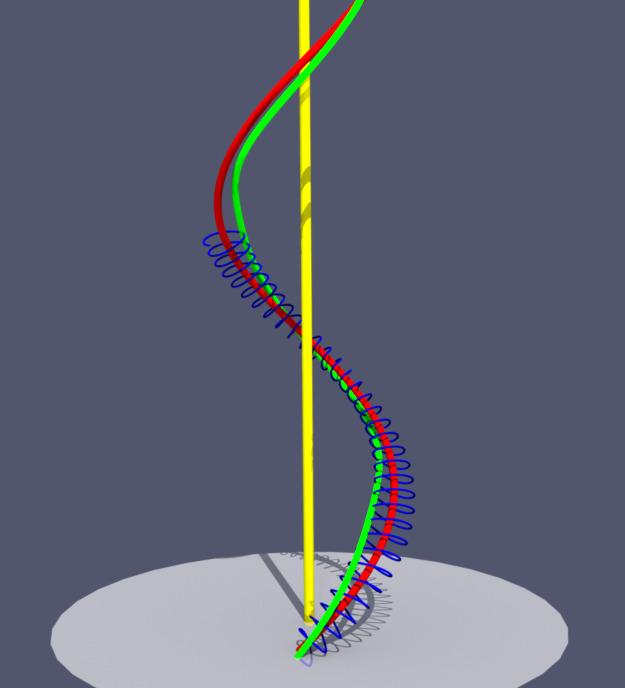}
	}}}
	\caption{Snapshot of a field line (in green) winding  around a steady vortex whose central field (in red) goes around the $z$ axis (in yellow) at a distance $r_0$ and a constant positive pitch $k$. In blue, a sample trajectory ${\bf u}_\perp$ of a point of the green field line.} 
	\label{fig_example_field_and_stream_lines}
\end{figure}

The 3D structure in the vicinity of the vortex defined by 
the set of parameters in Table \ref{tab_vortex_param} 
is shown in Figure \ref{fig_example_field_and_stream_lines}.
The magnetic field lines have been obtained by numerically integrating (\ref{eq_field_lines}) using the first order approximation components in (\ref{eq_B_first_order}). The same procedure has been used to construct the  streamlines of $\bf u$ and $\bf u_{\perp}$. The static, red field line traces the center of the vortex, while the green field line swirls around the vortex center. The blue line shows the trajectory of a point on the green line. The trajectory is ascending, indicating that the field line is moving upwards and thus carries an upwards directed Poynting flux. Note that there is no net Poynting flux neither towards nor away from the vortex center, a logical consequence of the steadiness of the system. 

\section{Example of a Uranus at solstice type tail vortex\label{sec_uranus}}

In this section, we verify that the close vicinity of the vortices which develop in the magnetic tail of a Uranus at solstice type can be conveniently described by the linear vortex model previously exposed. 

\begin{figure}
	\center{\rotatebox{0}{\resizebox{1\textwidth}{!}
			{\includegraphics{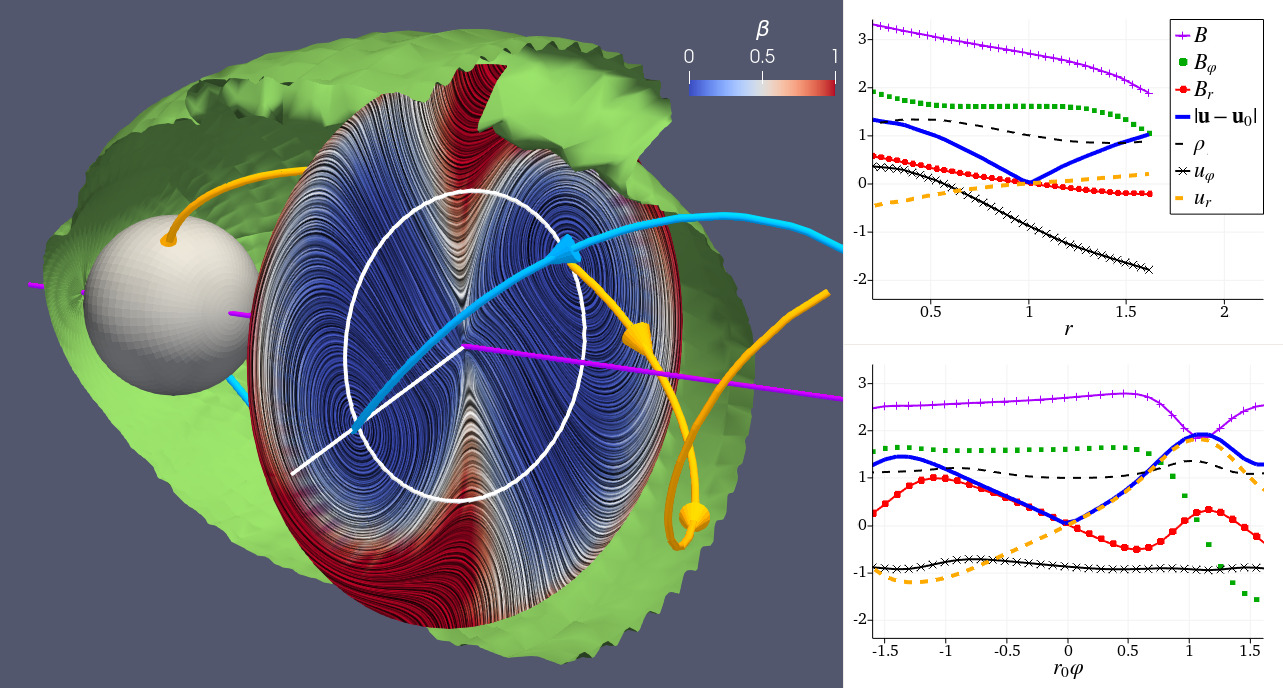}
	}}}
	\caption{Left panel: 3D MHD simulation of the magnetosphere of a rotating planet with a dipolar intrinsic magnetic field. The dipole is oriented perpendicularly with respect to the rotation axis (the magenta $z$ axis) while the surrounding unmagnetized supersonic stellar flows parallel to the rotation axis, from left to right. Such a system is stationary in the rotating frame, with two vortices winding around the rotation axis. The magnetic field lines located at the center of each vortex are shown in blue and yellow. 
In the plane perpendicular to the rotation axis, 
the streamlines of ${\bf u}_\perp$ show the motion of the magnetic field lines. The color scale is for the thermal pressure to magnetic pressure ratio $\beta$. The greenish surface is defined  by $\beta=2$ and loosely separates the high $\beta$ stellar wind plasma from the low $\beta$ planetary plasma. The white curves show the $r$ and the $\varphi$ lines crossing at the center of the blue vortex.	
Right panels: selected profiles along the white $r$ and $\varphi$ lines shown in the left}   
	\label{fig_3D_Uranus}
\end{figure}
We use the 3D MHD code MPI-AMRVAC to simulate the interaction of a stellar wind with a rotating magnetized planet. The code has been used with a general set-up similar to the one presented in \cite{Griton_etal_2018} except that the stellar wind is unmagnetized. The planet's rotation axis is parallel to the steady stellar wind flow and perpendicular to the planetary dipole, a situation reminiscent of Uranus at solstice time. 
The consequence of the absence of magnetic field in the stellar wind, as for the Uranus simulation in \cite{Toth_etal_2004}, implies that the system is steady in the rotating frame, which is a condition for the strict applicability of the above model. 
Without entering into the details of the simulation, which are of no particular interest in this place, the selected stellar wind parameters are typical of the conditions at Uranus with a wind velocity of 400\,km/s, a sonic Mach number $M=20$. In order to emphasize the effects of rotation, the rotation period has been set to  $\tau=1.76\,{\rm h}$, which is roughly ten times shorter than the effective period. The planetary dipole of intensity $D=11.893\,\mu {\rm T}R_{\rm U}^3$ (here $R_{\rm U}$ is the radius of Uranus) is located at the center of the planet.    

\begin{figure}
	\center{\rotatebox{0}{\resizebox{0.9\textwidth}{!}
			{\includegraphics{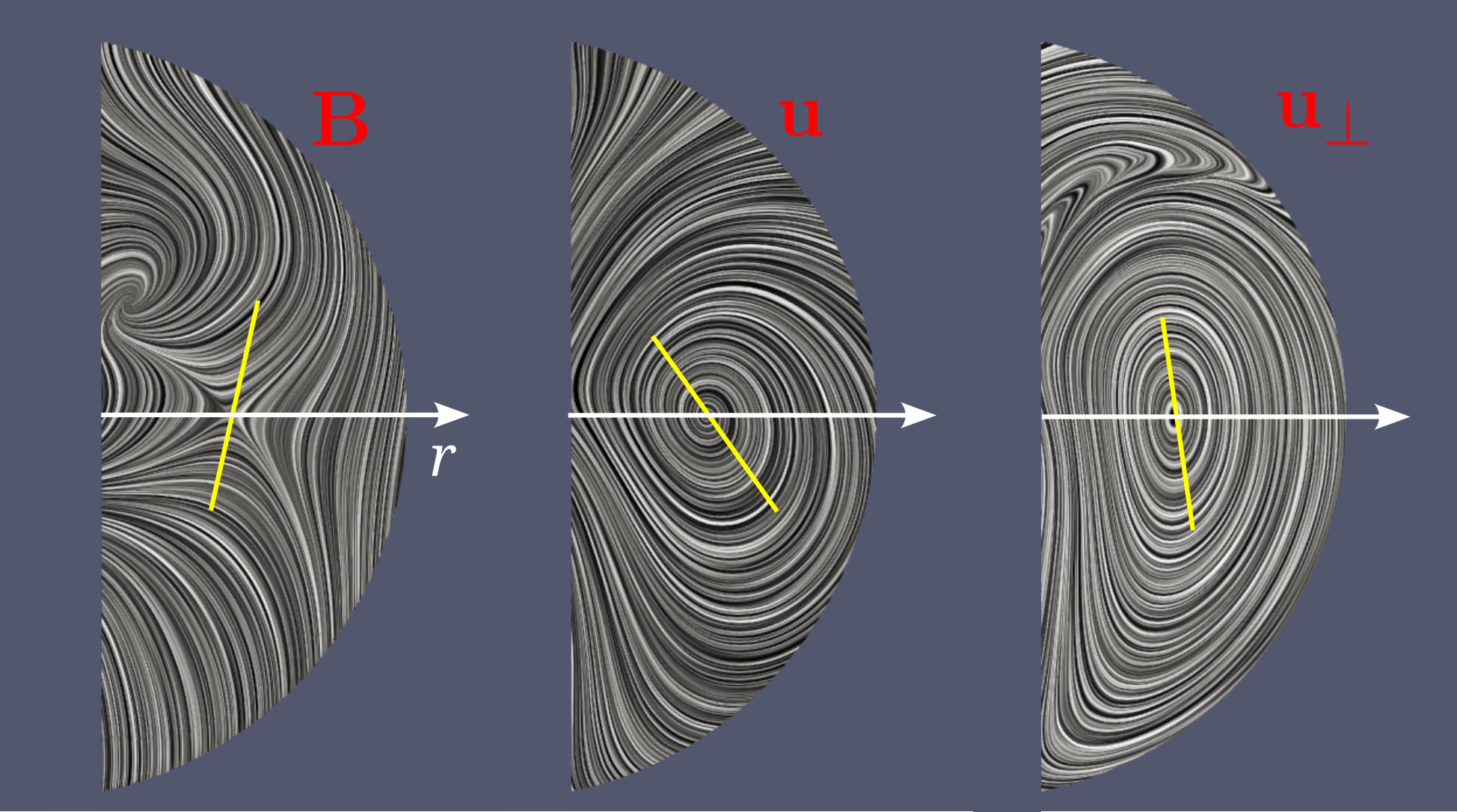}
	}}}
	\caption{Streamlines in the $(r,\xi)$ representation in the plane shown in the Uranus type simulation of Figure \ref{fig_3D_Uranus}. Only one vortex is shown here. The yellow segments indicate the orientation of  the corresponding conic section given in Table \ref{tab_vortex_structure_Uranus}.} 
	\label{fig_streamlines_Uranus}
\end{figure}

A snapshot of the simulation, after reaching steadiness in the rotating frame, is presented in Figure \ref{fig_3D_Uranus}.  
The figure is detailed enough to allow for a rough estimate of the model parameters for the vortex surrounding the blue field line. The objective is to determine the parameters of the vortex by completing a table similar to  Table  \ref{tab_vortex_param}. Hence, the normalized angular pitch of the vortex is easily found to be $kr_0=B_{\varphi 0}/B_{z0}\simeq-0.756$. (note that $B_{z0} <0$ and $B_{\varphi 0} >0$).     
The parameter $a$ can be estimated using its definition (\ref{eq_par_a}) and the $B_r$ profile in the bottom right panel of the figure from where we obtain $a=B_{z0}^{-1}\partial B_r/\partial\varphi\simeq 0.52$. The parameter $b\propto \partial B_\varphi/\partial \varphi$ is certainly small given the flatness of the $B_\varphi$ profile in the bottom right panel. We may better use $\partial B_r/\partial r=\tilde{A}_{10}-\lambda E_1=-bB_{z0}(1+k^2)$ and the $B_r$ profile from the upper right panel to compute $b=-\partial_r B_r/[B_{z0}(1+k^2)]\simeq -0.154$. The upper right panel also shows that $u_r$ and $B_r$ vary in antiphase, thus pointing to  $P=-1$ for the polarization parameter.  
Finally, observing that $\partial B_\varphi/\partial r \simeq 0$, we conclude that $\tilde{A}_{20}=\lambda E_{2}$ (see equation (\ref{eq_B_first_order})), from where we determine the remaining parameter $\tilde{A}_{10}=\tilde{A}_{20}\,a/b-bB_{z0}\,(1+k^2)\simeq 3.38$. 
All primary parameters for the vortex in the MHD simulation are summarized in Table \ref{tab_vortex_param_Uranus}     
\begin{table}[ht]
		\begin{tabular}{c|c|c|c|c|c|c|c}
			$r_0\Omega$ & $r_0k$ & $B_{\varphi 0}$ & $u_{\varphi 0}$ & $a$ &  $b$ & $r_0\tilde{A}_{10}$ & $P$ \\
			\hline\rule[0ex]{0pt}{2.5ex}
			$1.73$& $-0.756$ & $1.60$ & $-0.885$ & $0.52$ & $-0.154$ & $3.38$ & $-1$\\
		\end{tabular}
		\caption{Parameters for the vortices in the Uranus simulation.}
		\label{tab_vortex_param_Uranus}
\end{table}
and the derived parameters in Table \ref{tab_vortex_structure_Uranus}.   
\begin{table}[ht]
		\begin{tabular}{c|c|c|c|c|c|c|c|c}
			& $D_i$ &$\theta$ & $\mathcal{L}_1/\mathcal{L}_2$ & $s_1$ & $r_0\lambda$ &$\Lambda$& Ro & $v_{{\rm  ph},z}$\\
			\hline\rule[0ex]{0pt}{2.5ex}
			$B$ & $1.34$ &$77.3^\circ$ & $-$ &  $-$ & $-3.52$&$0.28$&$0.51$& $2.29$\\
			\hline\rule[0ex]{0pt}{2.5ex}
			$u$ & $-1.35$ &$125.4^\circ$& $1.55$ &  $-7.9$&$''$&$''$& $''$& $''$\\
			\hline\rule[0ex]{0pt}{2.5ex}
			$u_{\perp}$ & $-0.93$ &$98.2^\circ$& $2.10$ &  $-$ &$''$&$''$& $''$& $''$\\	
		\end{tabular}
		\caption{Derived quantities for the vortices in the  Uranus simulation, using the parameters from Table  \ref{tab_vortex_param_Uranus}.}
		\label{tab_vortex_structure_Uranus}
\end{table}
From (\ref{eq_A20_tilde}), in the non-rotating frame, the phase speed of the vortex along the $z$ axis is  
\begin{equation}
	v_{{\rm ph},z}=-\frac{\Omega}{k}= u_{z0}\pm B_{z0} \Lambda^{1/2}
	\label{eq_v_phz}.
\end{equation}
For $\Lambda=1$ (i.e. for a vanishing magnetic pressure gradient in the radial direction) reduces to equation (4) of  \cite{Pantellini_2020}. For the vortex in the MHD simulation, values in table \ref{tab_vortex_param_Uranus} give a phase speed $v_{{\rm ph},z}=-\Omega/k=2.29$, and imply the "$-$" sign in (\ref{eq_v_phz}). 
Hence, contrary to the example of the vortex in Table \ref{tab_vortex_structure}, the discriminant $D_B$ is now positive, indicating that the magnetic field lines in the $(r,\xi)$ representation are hyperbolic rather than elliptical. This is confirmed by the structure of the magnetic streamlines in the left panel of  Figure \ref{fig_streamlines_Uranus}. 
Yet, the discriminants $D_u$ and $D_{u\perp}$ being both negative, the corresponding streamlines in Figure \ref{fig_streamlines_Uranus} have an elliptical topology. The orientation of the axis (yellow segments) are in good qualitative agreement with the estimates given in Table \ref{tab_vortex_structure_Uranus}. The fact that the streamlines for $\bf u$ are not closed, but spiraling away from the vortex center (orientation of fluid motion is clockwise), is due to the vortices distance to the $z$ axis slowly increasing tailwards of the planet so that the system is not rigorously invariant along any 3D path with $r={\rm const}$ and $\varphi=kz$. 
On the other hand, the streamlines of ${\bf u}_\perp$ are closed. A consequence of the winding distance being much shorter for ${\bf u}_\perp$ than for ${\bf u}$ (see example of Fig. \ref{fig_example_field_and_stream_lines}). 

\begin{figure}
	\center{\rotatebox{0}{\resizebox{0.7\textwidth}{!}
			{\includegraphics{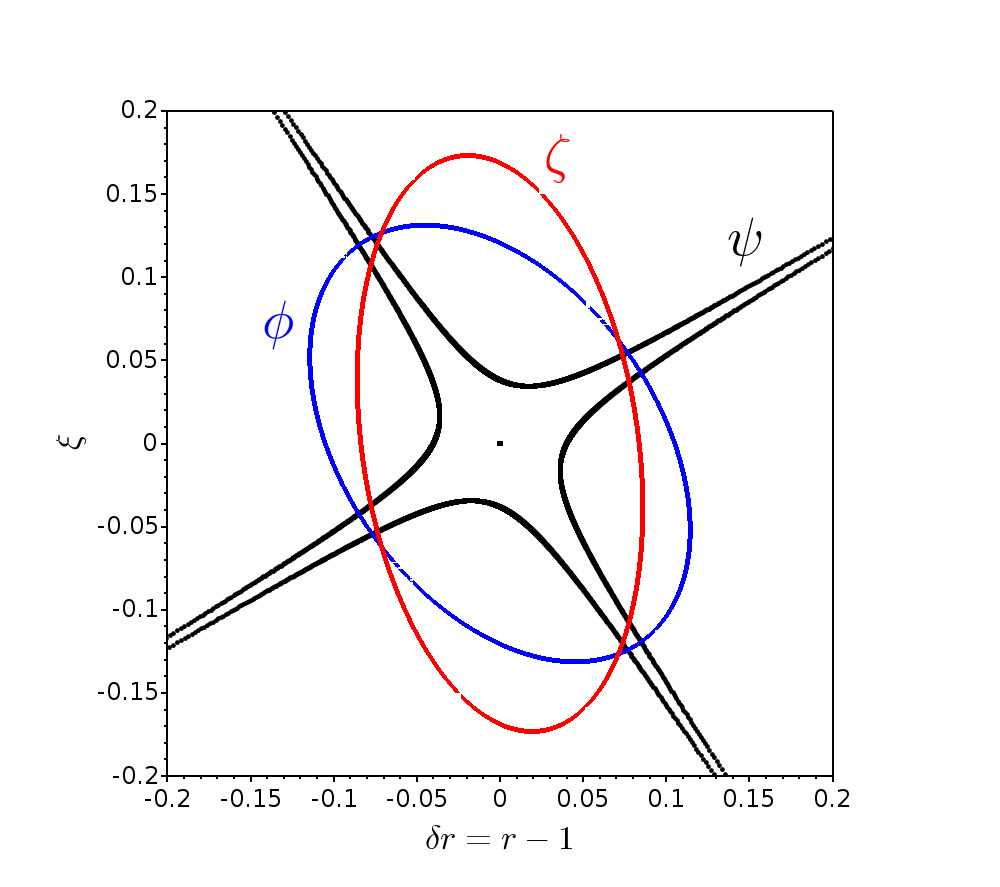}
	}}}
	\caption{Streamlines $\psi$, $\phi$ and $\zeta$ corresponding to 
 $\bf B$, $\bf u$ and ${\bf u}_\perp$, respectively. The parameters of the vortex, listed in Table \ref{tab_vortex_param_Uranus} have been deduced from the Uranus type simulation of Figure \ref{fig_3D_Uranus}.}
 \label{fig_streamlines_Uranus_model}
\end{figure}

We conclude by observing that, despite the fact that the simulated system does not rigorously satisfy all the constraints of the model, such as the $\xi$ invariance and the hypothesis of a uniform plasma density, 
there is good qualitative agreement between the streamlines observed in the 3D MHD simulation (see Figure \ref{fig_streamlines_Uranus}) and theoretical streamlines   shown in Figure \ref{fig_streamlines_Uranus_model} obtained by applying the linear model with the parameters of Table \ref{tab_vortex_param_Uranus}. 

\section{Conclusion\label{sec_conclusions}}
Vortex flows are ubiquitous in astrophysical plasmas.  
In magnetized plasmas they form Alfvénic structures which are    
conveniently described in the frame of incompressible, ideal MHD. Within this theoretical frame, and inspired by the vortex flows observed in numerical simulations of the magnetic tail of Uranus at solstice (see e.g. Figure \ref{fig_3D_Uranus}), we have presented a linear model for the structure of two identical, infinitely long, interlaced vortices of opposite polarity as shown in the idealized configuration of Figure \ref{fig_vortex_geometry}). Each vortex is an Alfvén mode, whose linear structure is ultimately imposed by the conditions in the surrounding plasma, allowing for a large variety of flows, even in the case of an idealized configuration.  Assuming a polytropic and constant density plasma, it turns out that eight parameters are generally necessary to fully describe the flow near the vortex center (see (\ref{eq_problem_parameters})). 
In Section \ref{sec_uranus}, the linear model has been shown to 
conveniently described the vortices which develop in a full 3D MHD simulation of the magnetosphere of a Uranus like planet.  
One may be tempted to consider the case of vortices winding around the $z$ at a varying distance or with a varying pitch. This will make the model more general, but probably too complex to be of any practical utility. The model may be extended to include thermal pressure effects, which could be of some interest in high beta plasmas.  

\section*{Acknowledgments} 
This work has been financially supported by the PLAS@PAR project and by the National Institute of Sciences of the Universe (INSU).
Thank you to the anonymous reviewer for valuable comments which helped to improve the paper and to ease the task of the reader.   

\begin{appendices}

\section{Explicit form of the matrices of the differential equation system\label{app_matrices}}

Recalling the correspondence between the components of the 6 dimensional vector $\bf X$ and the components of the magnetic field vector $\bf B$
and the velocity vector $\bf u$
\begin{equation}
{\bf X}\equiv\begin{pmatrix}
	X_1\\
	X_2\\
	X_3\\
	X_4\\
	X_5\\
	X_6 \end{pmatrix}
 \equiv  
 \begin{pmatrix}
	B_r\\
	B_\varphi\\
	B_z\\
	u_r\\
	u_\varphi\\
	u_z\end{pmatrix}
\end{equation}
the $6\times6$ matrices in the system of differential equations (\ref{eq_diff}) 
take the form
\begin{equation}
M_r=r\begin{pmatrix}{{X_4}} & 0 & 0 & -{{X_1}} & 0 & 0\\
	0 & {{X_4}} & 0 & 0 & -{{X_1}} & 0\\
	0 & 0 & {{X_4}} & 0 & 0 & -{{X_1}}\\
	0 & {{X_2}} & {{X_3}} &  \, {{X_4}} & 0 & 0\\
	0 & -{{X_1}} & 0 & 0 &  \, {{X_4}} & 0\\
	0 & 0 & -{{X_1}} & 0 & 0 &  \, {{X_4}}\end{pmatrix}\mbox{}
\end{equation}
and
\begin{equation} 
	M_\xi=\begin{pmatrix}{{X_5}}-kr\, {{X_6}} & 0 & 0 & kr\, {{X_3}}-{{X_2}} & 0 & 0\\
		0 & {{X_5}}-kr\, {{X_6}} & 0 & 0 & kr\, {{X_3}}-{{X_2}} & 0\\
		0 & 0 & {{X_5}}-kr\, {{X_6}} & 0 & 0 & kr\, {{X_3}}-{{X_2}}\\
		kr\, {{X_3}}-{{X_2}} & 0 & 0 &  \, {{X_5}}-kr \, {{X_6}} & 0 & 0\\
		{{X_1}} & kr\, {{X_3}} & {{X_3}} & 0 &  \, {{X_5}}-kr  \, {{X_6}} & 0\\
		-kr\, {{X_1}} & -kr\, {{X_2}} & -{{X_2}} & 0 & 0 &  \, {{X_5}}-kr \, {{X_6}}\end{pmatrix}\mbox{}.
\end{equation}
As the inverse of $M_r$ is required to compute the matrix 
$\tilde{M}_\xi=M_r^{-1} M_\xi$ and the vector $\tilde{A}_\xi=M_r^{-1} A_\xi$ which appear in the  normal form of the system of differential equations (\ref{eq_diff_normal}), we also give its explicit form hereafter:  
\begin{eqnarray} &&M_r^{-1}=\frac{1}{r X_4(X_4^2 -X_1^2)}\times \nonumber\\
	&&\begin{pmatrix}
		X_4^{2}-X_1^2 & -X_1 X_2 & -X_1 X_3 &  
		(X_4^{2}-X_1^2)X_1/X_4  & -X_1^{2} X_2/X_4 & -X_1^{2} X_3/X_4\\
		0 & X_4^2 & 0 & 0 & X_1 X_4 & 0\\
		0 & 0 & X_4^2 & 0 & 0 & X_1 X_4\\
		0 & -X_2 X_4 & -X_3 X_4 & X_4^{2}-X_1^2 & -X_1 X_2 & -X_1 X_3\\
		0 & X_1 X_4 & 0 & 0 & X_4^2 & 0\\
		0 & 0 & X_1 X_4 & 0 & 0 & X_4^2
	\end{pmatrix}.
\end{eqnarray}

\end{appendices}

\bibliography{vortex.bib}

\end{document}